\documentclass[preprint,showpacs,preprintnumbers,amsmath,amssymb,nofootinbib]{revtex4}

\usepackage{graphicx}
\usepackage{color}
\usepackage{dcolumn}
\usepackage{bm}

\begin{document}


\title{Cosmology in a certain vector-tensor theory of gravitation}

\author{Roberto Dale}
\email{rdale@umh.es}  
\affiliation{
Departamento de Estad\'{\i}sica, Matem\'atica e Inform\'atica, 
Universidad Miguel Hernandez, 03202 Elche, Alicante, Spain    \\
}%

\author{Diego S\'aez}
\email{diego.saez@uv.es}
\affiliation{Departamento de Astronom\'{\i}a y Astrof\'{\i}sica, Universidad de Valencia,
46100 Burjassot, Valencia, Spain\\
}%

\date{\today}

\begin{abstract}
We study relevant cosmological topics 
in the framework of a certain vector-tensor theory of gravitation
(hereafter VT).
This theory is first compared with the  
so-called extended electromagnetism (EE). 
These theories have a notable resemblance and both explain 
the existence of a cosmological constant.
It is shown that, in EE, a positive dark energy density requires a
Lagrangian leading to quantum ghosts, whereas VT
is free from these ghosts. On account of this fact, 
the remainder of the paper is devoted to study cosmology in the 
framework of VT. 
Initial conditions, 
at high redshift, are used to solve the evolution equations of all the VT scalar modes. 
In particular, a certain scalar mode characteristic of VT --which does not appear in 
general relativity (GR)-- 
is chosen in such a way that it evolves separately. In other words, the scalar  
modes of the standard model based on GR do not affect the evolution of the 
VT characteristic mode; however, this scalar mode influences the evolution of the standard 
GR ones. Some well known suitable codes (CMBFAST and COSMOMC) have been modified 
to include our VT initial conditions and evolution equations, which are fully general.
One of the resulting codes --based on standard statistical methods-- has been used 
to fit VT predictions and observational evidences about both Ia supernovae and 
cosmic microwave background anisotropy. 
Seven free parameters are used in this fit. Six of them are often used 
in GR cosmology and the seventh one is characteristic of VT. 
From the statistical analysis it follows that VT seems to be advantageous against GR in 
order to explain cosmological observational evidences.
\end{abstract}



\maketitle

\section{Introduction}
\label{sec:1}

Extended electromagnetism (EE) was proposed in paper \cite{bm091}.
The basic fields of this theory are the metric $g^{\mu \nu} $
and the electromagnetic field $A^{\mu} $. The fundamental symmetry is
$A^{\mu}  \rightarrow  A^{\mu} + \nabla^{\mu} \Phi$, 
with $\nabla_{\mu} \nabla^{\mu} \Phi =0$; which is different from the 
standard U(1) gauge symmetry. 

Some cosmological applications of EE were 
discussed in various papers \citep{bm092,bm011,dal12}.
In Dale \& S\'aez \cite{dal12}, the variational formulation of EE
was revisited, and the cosmological linear perturbations
were studied by using the well known Bardeen formalism 
\citep{bar80,huw97}. 

There is a vector-tensor (VT) theory of gravitation,
studied in \cite{dal09}, which has a notable resemblance with EE.
The post-Newtonian parametrized limit of VT is identical to that of 
general relativity (GR).
Moreover, this theory was proved to be viable in \cite{bm093}
(see below for more details).
Here, the theories EE and VT are compared 
to conclude that, although they 
give the same results in cosmology,
there are some problems with EE quantification. On account of these facts, our cosmological results 
are presented in the framework of VT.

The initial conditions for the evolution of scalar perturbations 
are taken in the radiation dominated era, at redshift 
$z \sim 10^{8} $, when the perturbations of cosmological interest
are outside the effective horizon (see paper \cite{mb95} for details).
By using these initial conditions, the linear equations satisfied by the 
scalar perturbations 
are numerically solved, and the cosmic microwave background (CMB) anisotropy is 
estimated. Since all the scalar perturbations are evolved from 
the radiation dominated era, it may be seen how metric perturbations
gradually deviate from 
the GR ones. Deviations arises at some redshift to be
numerically estimated, without {\it a priori} assumptions about its 
possible value.

There are well tested codes which are ready to do
some calculations (evolution of scalar perturbations, CMB analysis, and so on) 
for standard cosmological models based on GR; e.g., CMBFAST \citep{seza96} and CAMB \citep{lew00}. 
These codes may be modified
to work in the framework of VT (or EE). In spite of the fact that CMBFAST 
is not currently maintained, its last version is good enough 
for us and, moreover, its equations are essentially written 
by using the Bardeen 
formalism in the version of \cite{mb95}, which is the same formalism used 
to study VT along this paper. By this reason, 
we may easily modify CMBFAST to describe 
cosmological evolution in VT.
The necessary modifications --based on references \cite{bar80}, 
\cite{mb95}, and \cite{dal12}-- are lengthy but straightforward.
The code COSMOMC \citep{lew02} has been also
modified for statistical analysis in VT;
namely, to fit theoretical predictions and observations by using
a set of parameters (see next sections for details).

Our signature is (--,+,+,+). Greek (Latin) indices run from $0$ to $3$ 
(1 to 3). The symbol $\nabla $ ($\partial $) stands
for a covariant (partial) derivative. The antisymmetric tensor $F_{\mu \nu} $
is defined by the relation 
$F_{\mu \nu} = \nabla_{\mu} A_{\nu } - \nabla_{\nu} A_{\mu }$,
where $A^{\mu }$ is the vector field of the theory under consideration (EE or VT).
Quantities $R_{\mu \nu}$, $R$, and $g$ are the covariant components of the Ricci 
tensor, the scalar curvature and
the determinant  of the matrix $g_{\mu \nu}$ formed by the covariant components 
of the metric, respectively. The gravitational constant
is denoted $G$. Units are chosen in such a way that the speed of
light is $c=1$. The scale factor is $a$. In flat universes,
the present value of $a$ is arbitrary. We take a=1. The coordinate and conformal
times are $t$ and $\tau $, respectively.
Whatever quantity $D$ may be, $D_{B} $  stands for
its background value and $\dot{D}$ is its derivative with respect to 
the conformal time.
 
This paper is structured as follows. In Sec. \ref{sec:2}, some general aspects of VT and EE and 
the cosmological background equations of both theories are presented and compared. 
In Sec. \ref{sec:3},
the evolution equations of all the VT cosmological scalar modes and the initial 
conditions necessary to their numerical integration are found. Numerical results 
are obtained with our modified versions of CMBFAST and COSMOMC. These results  are analyzed in 
Sec. \ref{sec:4} and, finally, Sec. \ref{sec:5} is a general discussion 
about methodology and 
conclusions.

\section{Basic equations of EE and VT. Background universe}
\label{sec:2}   
  
The basic equations of EE may be derived from the
action \citep{dal12}: 
\begin{equation}
I = \int \left[ \frac {R} {16\pi G}
- \frac {1}{4} F^{\mu \nu } F_{\mu \nu }
+\gamma (\nabla_\mu A^{\mu})^{2}   
+ J^{\mu} A_{\mu} - \rho (1 + \epsilon)
\right] \,\sqrt { - g} \,d^4 x  \ ,
\label{1.1}
\end{equation}   
where $\gamma $ is a dimensionless arbitrary parameter,
$J^{\mu} $ is the electrical current, 
$\rho $ is the conserved energy density of an isentropic perfect
fluid, and $\epsilon$ is its internal energy density 
[see papers \cite{dal12} and \cite{haw99} 
for details].
  
From action (\ref{1.1}), we have found two coupled field
equations. The first equation is a generalization of the
GR equation describing gravity. 
This equation may be written in the following form:
\begin{equation}  
G^{\mu \nu} = 8\pi G (T^{\mu \nu}_{GR} + T^{\mu \nu}_{EE}) \ ,
\label{fieles}
\end{equation} 
where $G^{\mu \nu}$ is the Einstein tensor, $T^{\mu \nu}_{GR}$ is the energy momentum 
tensor of a fluid as it appears in GR, and the energy momentum tensor of 
the electromagnetic field --in EE theory-- is $T^{\mu \nu}_{EE}$.
The second equation is a generalization 
of Maxwell equation in curved space-time. This equation reads as follows:
\begin{equation}
\nabla^{\nu} F_{\mu \nu} = J_{\mu} + J^{^{A}}_{\mu} \ ,
\label{1.3}
\end{equation} 
where $J^{^{A}}_{\mu} = -2 \gamma \nabla_{\mu} (\nabla \cdot A)$ 
--with  $\nabla \cdot A = \nabla_{\mu} A^{\mu} $-- plays the role of a 
new fictitious current. From this last equation one easily finds the relation
\begin{equation}
\nabla^{\mu} (J_{\mu} + J^{^{A}}_{\mu})=0 
\label{conlaw}
\end{equation} 
and, consequently,  
the total current $J_{\mu} + J^{^{A}}_{\mu}$ is
conserved.

The energy momentum tensors $T^{\mu \nu}_{GR}$ and $T^{\mu \nu}_{EE} $
involved in Eqs.(\ref{fieles}) are 
\begin{equation}
T^{\mu \nu}_{GR} = (\mu + P)U^{\mu}U^{\nu} + Pg^{\mu \nu}  
\label{emtrg}
\end{equation}
\begin{eqnarray}
T^{\mu \nu}_{EE} &=& F^{\mu}_{\,\,\,\, \alpha}F^{\nu \alpha}
- \frac {1}{4} g^{\mu \nu} F_{\alpha \beta} F^{\alpha \beta} \nonumber \\ 
& &
+2\gamma [ \{A^{\alpha}\nabla_{\alpha} (\nabla \cdot A) + \frac {1}{2}(\nabla \cdot A)^{2}\}
g^{\mu \nu}-
A^{\mu}\nabla^{\nu} (\nabla \cdot A) - A^{\nu}\nabla^{\mu} (\nabla \cdot A)
] \ .
\label{emtee}
\end{eqnarray}  
The part of this last energy momentum tensor depending on $\gamma $ appears 
in EE but not in Einstein-Maxwell (E-M) theory. The two first terms of 
this tensor also appear in E-M.

In vector-tensor theories of gravitation,
there are also two fields, the metric $g_{\mu \nu}$, and a four-vector $A^{\mu}$
which has nothing to do with the electromagnetic field.
Various of these theories have been developed
(see \cite{wil93}, \cite{wil06} and references cited there). They 
are based on the general action \citep{wil93}:

\begin{eqnarray}
I &=& \int [ R/16\pi G + \omega A_\mu  A^\mu  R
+ \eta R_{\mu \nu }
A^\mu  A^\nu  - 
\varepsilon F_{\mu \nu } F^{\mu \nu }
+\gamma \,\nabla_\nu  A_\mu  \nabla^\nu  A^\mu  \nonumber \\ 
& &
- \rho (1 + \epsilon)]
\sqrt { - g} \,d^4 x  \ ,
\label{VT.1}
\end{eqnarray}   
where $\omega$, $\eta $, $\varepsilon$,
and $\gamma$ are arbitrary parameters.
The tensor $F_{\mu \nu} $ --defined above--
is not the electromagnetic one.
In action (\ref{VT.1}), it is implicitly assumed that
there are no couplings of $A_{\mu}$ with matter fields and 
electrical currents.

We are interest in the theory VT, which may be derived from the action 
(\ref{VT.1}) for $\omega=0 $, and $\eta = \gamma $ (see 
\cite{dal09}). With these parameters, the 
Lagrangian of Eq.~(\ref{VT.1}) is easily proved to be 
equivalent to that involved in the following action
(the difference is a total divergence): 
\begin{equation}
I = \int \left[ \frac {R} {16\pi G} 
+ (\frac {\gamma}{2} - \varepsilon ) F^{\mu \nu } F_{\mu \nu } 
+\gamma \, (\nabla_\mu A^{\mu})^{2}
- \rho (1 + \epsilon)
\right] \,\sqrt { - g} \,d^4 x  \ .
\label{VT.2}
\end{equation}  
A complete discussion about 
ghosts and unstable modes in VT was presented in section 3.1.3 of reference \cite{bm093}, 
where it was proved  
that there are no problems with 
this theory for $2\varepsilon - \gamma > 0$. This condition is hereafter assumed. 

Let us now compare actions (\ref{1.1}) and (\ref{VT.2}). 
Action (\ref{VT.2}) does not contain the term 
$J^{\mu} A_{\mu} $ involved in Eq.~(\ref{1.1}), and 
the coefficient of 
$F^{\mu \nu } F_{\mu \nu } $ is arbitrary in action (\ref{VT.2})
whereas it takes on the value $-1/4 $ in Eq.~(\ref{1.1}). 
These differences are consistent with the fact that EE is 
a theory of electromagnetism whereas VT is a theory of gravitation,
in which, $A^{\mu} $ and $F_{\mu \nu}$
have nothing to do with electromagnetism.

The fundamental equations of VT are easily obtained from the 
action (\ref{VT.2}). They have the following form:
\begin{equation}  
G^{\mu \nu} = 8\pi G (T^{\mu \nu}_{GR} + T^{\mu \nu}_{VT}) \ ,
\label{fieles_vt}
\end{equation} 
\begin{equation}
2(2\varepsilon - \gamma)\nabla^{\nu} F_{\mu \nu} = J^{^{A}}_{\mu} \ ,
\label{1.3_vt}
\end{equation} 
with
\begin{eqnarray}
T^{\mu \nu}_{VT} &=& 2(2\varepsilon - \gamma) [F^{\mu}_{\,\,\,\, \alpha}F^{\nu \alpha}
- \frac {1}{4} g^{\mu \nu} F_{\alpha \beta} F^{\alpha \beta}] \nonumber \\ 
& &
-2\gamma [ \{A^{\alpha}\nabla_{\alpha} (\nabla \cdot A) + \frac {1}{2}(\nabla \cdot A)^{2}\}
g^{\mu \nu}-
A^{\mu}\nabla^{\nu} (\nabla \cdot A) - A^{\nu}\nabla^{\mu} (\nabla \cdot A)
] \ .
\label{emtee_vt}
\end{eqnarray}  
From Eq.~(\ref{1.3_vt}) one easily gets the relation 
\begin{equation}
\nabla^{\mu} J^{^{A}}_{\mu} = 0 \ , 
\label{confic}
\end{equation}
which may be seen as the conservation law of the fictitious current $J^{^{A}}_{\mu}$
defined above.

Let us finally answer the following question: Why the last terms 
of Eqs.~(\ref{emtee}) and~(\ref{emtee_vt}) have the same form 
but opposite signs?. The answer to this question may be found
in a previous paper \citep{dal12}, where we presented an exhaustive variational 
formulation of EE based on the action (\ref{1.1}). 
Some important aspects of this variational formulation are pointed out here,
with the essential aim of answering the above question. 
Our variational method is described in \cite{haw99} (see section 3.3), 
where it is used to study the evolution of an isentropic fluid 
satisfying a certain conservation law. 
The same method may be easily generalized to deal with action (\ref{1.1}),
in which  the vector field $A^{\mu}$,
the fluid four-velocity $U^{\mu}$, and the metric $g_{\mu \nu} $,
must be successively varied.

The field $A^{\mu}$ is first varied --for arbitrary $g_{\mu \nu} $ and $U^{\mu}$--
to get Eqs.~(\ref{1.3}) and the conservation law (\ref{conlaw}), which may be 
rewritten as follows: $\partial_{\mu} [\sqrt{-g} (J^{\mu} + J^{A\mu})]=0$. 

In a second step, only the four-velocity is varied, whereas the field $A^{\mu} $ is 
any arbitrary solution of Eqs.~(\ref{1.3})--(\ref{conlaw}) for 
arbitrary metric. Then, the charge density $\rho_{q}$ is adjusted to keep the total 
current $ J^{\mu} + J^{A\mu} $ conserved ($ J^{\mu} = \rho_{q} U^{\mu}$). 
Thus, the following equation is obtained 
\begin{equation}
(\mu + P)U^{\mu}\nabla_{\mu}U^{\nu} = -\nabla_{\mu}P(g^{\mu \nu}+U^{\mu}U^{\nu})+
F^{\mu \nu}J_{\mu}+(\nabla^{\mu}J^{^{A}}_{\mu})A^{\nu} \ .
\label{1.5}   
\end{equation}
This equation was already derived in \cite{dal12}. See also \cite{haw99} 
for similar calculations in GR.

Finally, only the metric is varied, whereas the vector
$A^{\mu}$ is fixed as in the second step, and vector 
$V^{\mu} $ satisfies 
Eq.~(\ref{1.5}) whatever $g^{\mu \nu}$ may be; hence,
since density $\rho_{q}$ has been appropriately adjusted (see above),
the conservation law 
$\partial_{\mu} [\sqrt{-g} (J^{\mu} + J^{A\mu})] =0$ is satisfied along 
the flow lines for arbitrary $g^{\mu \nu}$. Hence, 
$\sqrt{-g} (J^{\mu} + J^{A\mu})$ is unchanged when the metric is varied (see \cite{haw99}); 
namely, we can write
\begin{equation} 
\Delta_{g} (\sqrt{-g} J^{\mu}) =  - \Delta_{g} (\sqrt{-g}J^{^{A}\mu})  \ ,
\label{crucial}
\end{equation} 
where $\Delta_{g} $ stands for a metric variation.

Equation~(\ref{crucial}) implies that the term $J^{\mu} A_{\mu} $ --involved in 
action~(\ref{1.1})-- is equivalent to $-J^{A\mu} A_{\mu} $ 
under the $\Delta_{g} $ variations necessary  to get the
energy-momentum tensor. Taking into account this fact,  
plus the identity $\nabla^{\mu}[A_{\mu} (\nabla \cdot A)] = (\nabla \cdot A)^{2} 
+ A_{\mu} \nabla^{\mu} (\nabla \cdot A) $, and the definition of the 
fictitious current $J^{A\mu} $, 
it is easily proved that the Lagrangian 
densities $\gamma (\nabla_\mu A^{\mu})^{2}                             
+ J^{\mu} A_{\mu}$ 
and $-\gamma (\nabla_\mu A^{\mu})^{2} $ are fully equivalent under $g_{\mu \nu}$ variations
(their difference is a total divergence); therefore, the energy-momentum 
tensors of EE and VT may be calculated from the Lagrangian densities 
$-\gamma (\nabla_\mu A^{\mu})^{2} $ and
$\gamma (\nabla_\mu A^{\mu})^{2} $, respectively. Hence,
the signs appearing in the last terms of Eqs.~(\ref{emtee}) and~(\ref{emtee_vt}) must be
opposite. This fact will play an important role later in this paper. It is a 
consequence of the conservation law~(\ref{conlaw}).

Let us now consider 
a flat uncharged homogeneous and isotropic background universe
with matter and radiation in both EE and VT. In this flat background, the
metric has the Robertson-Walker form. Moreover, the following relations are 
satisfied: $A_{i} =0$ and $F_{\mu \nu} =0$ \citep{dal12}.

In EE, 
Eqs.(\ref{fieles}), (\ref{emtrg}), and (\ref{emtee}) lead to
\begin{equation}
3 \frac {\dot{a}^{2}}{a^{2}}= 8 \pi G a^{2} (\rho_{B}+\rho^{A}_{B})
\label{baseq1}
\end{equation} 
\begin{equation}
-2 \frac {\ddot{a}}{a} + \frac {\dot{a}^{2}}{a^{2}} = 8 \pi G a^{2} (P_{B}+P^{A}_{B})
\label{baseq2}
\end{equation} 
where $\rho_{B} $ and $P_{B}$ are the background energy density and pressure 
of the cosmological fluid (baryons, dark matter, massless neutrinos and radiation),
and quantities  $\rho^{A}_{B} $ and $P^{A}_{B} $ are associated to the part of the 
energy-momentum tensor (\ref{emtee}) depending on $\gamma $.
From this part, the following relation is easily obtained:
\begin{equation}
\rho^{A}_{B} = -P^{A}_{B} = - \gamma (\nabla \cdot A)_{B}^{2} \ .
\label{eqest}
\end{equation}  
Hence, constant $\gamma $ must be negative to have a 
positive energy density.   

In VT, Eqs.~(\ref{baseq1}) and~(\ref{baseq2}) hold, but 
$\rho^{A}_{B} $ and $P^{A}_{B} $ must be defined by using the 
energy-momentum tensor (\ref{emtee_vt}); from which, one easily gets:
\begin{equation}
\rho^{A}_{B} = -P^{A}_{B} = \gamma (\nabla \cdot A)_{B}^{2} \ .
\label{eqest_vt}
\end{equation}  
Therefore, constant $\gamma $ must be positive to have
$\rho^{A}_{B} >0$.   

We have shown that simple applications to cosmology fix the sign 
of $\gamma $ in both EE and VT. This sign is irrelevant in cosmology,
but it is important in quantum field theory. On account of Eq.~(\ref{1.1}),
a negative sign of the coefficient $\gamma $ would lead to 
quantum ghosts. Hence, we hereafter develop our cosmological 
estimations in the framework of VT (with positive $\gamma $). 
In this theory, apart from the positive constant $\gamma $, 
whose value is unknown. There is a second 
constant $\varepsilon $ in Eq.~(\ref{VT.2}), which must satisfy 
the condition $\varepsilon > \frac {\gamma}{2} $, but the 
exact value of $\varepsilon $ keeps unknown.

Let us now study other background equations in VT. From Eq.~(\ref{1.3_vt}) one
easily gets:
\begin{equation}  
\Xi_{B} \equiv (\nabla \cdot A)_{B} = - \frac {1}{a^{2}} [ \dot{A}_{0B} 
+ 2 \frac {\dot{a}}{a} A_{0B}] = constant   \ ,
\label{coscons}
\end{equation} 
where $A_{0B} $ is the time component of $A_{\mu} $ in the background. Eq.~(\ref{coscons}) 
describes the evolution of this component. This equation may be numerically solved 
for appropriate initial conditions to get function $A_{0B} (\tau)$. 
From Eqs.~(\ref{eqest_vt}) and~(\ref{coscons}) one easily concludes that,
at zero order (in the background), 
the energy density of the field $A^{\mu} $ and its pressure have the same 
absolute value and opposite signs,
which means that $\rho^{A}_{B}$ plays the role of dark energy with the vacuum 
equation of state 
$W = P^{A}_{B}/\rho^{A}_{B} = -1$.

For vacuum energy ($W=-1$) and a flat background, CMBFAST uses 
Eqs.~(\ref{baseq1}) and~(\ref{baseq2}) with $\rho^{A}_{B}=\rho_{v}$, 
where $\rho_{v} $ is the vacuum energy; hence, in order
to modify CMBFAST for VT calculations, the CMBFAST background equations are 
valid, but the new Eq.~(\ref{coscons}) must be included.
According to Eq.~(\ref{eqest_vt}), in this new equation we set
\begin{equation}  
\Xi_{B} = S_{gn} \Big(\frac {\rho_{v}} {\gamma} \Big)^{1/2} \ ,
\label{xib}
\end{equation} 
where the value of $S_{gn} $ is either $+1$ or $-1$. The $S_{gn} $ value
fixes the arbitrary sign of $\Xi_{B}$.
In addition to $\rho_{v} $, we have the arbitrary parameters $\gamma $ and
$S_{gn} $. The integration of the new background equation
\begin{equation}
\dot{A}_{0B} = -a^{2} \Xi_{B} -2 \frac {\dot{a}}{a} A_{0B}
\label{newbeq}
\end{equation}
requires the initial value of $A_{0B}$, which is taken at the initial redshift $z_{in} = 10^{8}$ 
(as it is done 
for any variable). At this high redshift, during the radiation dominated era, there are
power law functions of $\tau $ satisfying the background field equations. In fact, it is easily verified that 
the following functions 
$a = \alpha \tau^{\zeta} $ and $A_{0B}  = \beta \tau^{\delta}$ 
satisfy the above background field equations --in the radiation dominated era--
for $\zeta=1 $, $\delta=3 $. Then, at $z_{in} $, one finds:
\begin{equation}
\tau_{in} = \Big( \frac {\dot{a}}{a} \Big)_{in}^{-1}\ , \,\,\,\,\,\, 
(A_{0B})_{in} = - \frac {\Xi_{B}}{5(1+z_{in})^{2}}  
\Big/ \Big( \frac {\dot{a}}{a} \Big)_{in}   
\ .
\label{inib0}
\end{equation}
Since CMBFAST rightly calculates the initial value of $\dot{a}/a$, 
the value of $A_{0B}$ at $z_{in} $ is not a free parameter. It is given in terms of 
$\rho_{v} $, $S_{gn} $ and $\gamma $ by Eqs.~(\ref{xib}) and~(\ref{inib0}).

After proving that $\Xi_{B} $ and $A_{0B}$ are both proportional to
$S_{gn}\gamma^{-1/2}$ [see Eqs.~(\ref{xib}) --~(\ref{inib0})], the 
background equations of VT might be easily solved 
for $\gamma=1$, $S_{gn}=+1$, and  for appropriate
amounts of baryons, dark matter, and photons 
--similar calculations were done by Dale \& S\'aez \cite{dal12} in EE--
nevertheless, 
massless neutrinos would require a more complicated treatment. These neutrinos are 
taken into account in CMBFAST and also in our modification of this code, 
in which we include
the VT parameters $\gamma$ and $S_{gn}$, the VT background equation 
(\ref{newbeq}) and, the new initial condition for $A_{0B}$ [see Eq.~(\ref{inib0})]. 
Any other aspect of the CMBFAST background evolution is not altered
at all.

\section{Cosmological scalar modes and initial conditions}  
\label{sec:3}

There are no tensor modes associated to the vector field $A^{\mu} $ and,
consequently, the evolution of tensor cosmological perturbations (primordial gravitational waves) 
is identical in GR and VT.

The vector modes involved in GR decrease as a result of expansion \citep{mor07}; hence, 
they are expected to be negligible at redshifts close to recombination and decoupling. 
Since significant 
vector modes might produce interesting effects 
\citep{mor07,mor08} at these low redshifts, 
it is interesting the study of vector-tensor theories, 
which include the vector modes of GR plus an additional one associated to the 
vector field $A^{\mu} $. The field equations --of the vector-tensor theory-- 
would couple all these modes which
could evolve in an appropriate way justifying the existence of non negligible 
vector modes at redshifts close to $1000$.
The study of vector modes in VT and also in other 
vector-tensor theories of gravitation is in progress. 

Since the effects of 
vector and tensor modes on the CMB are expected to be small,
this section is 
devoted to the study of scalar perturbations
in the framework of VT.

The code CMBFAST solves the evolution equations of the scalar modes 
in GR cosmologies. In the flat case, these equations are written in terms of
a certain set of scalar modes, whose initial values 
--at redshift $z=10^{8}$-- are appropriately obtained \citep{mb95}.
Calculations are performed in the synchronous gauge. 
In order to modify CMBFAST in the simplest way -for applications to VT--
the gauge and the scalar modes used in this code must be maintained, and
a new scalar mode associated to $A^{\mu} $
must be added. New terms depending on the new mode modify 
the CMBFAST equations (standard cosmology),
and a new equation for the evolution of the  
$A^{\mu} $ mode must be also added. Finally,
the initial values of all the coupled modes must be calculated 
at the chosen initial redshift.

For a
standard flat cosmological background in GR,
the formalism described in \cite{bar80} involves
the scalar perturbations associated to the metric, the four-velocity, and the
energy-momentum tensor of a cosmological fluid. These perturbations are expanded in terms of 
scalar harmonics as follows:
\begin{eqnarray}
& &
g_{00}=-a^{2}(1+2\tilde{A}Q^{(0)}), \,\,\,\, g_{0i} = -a^{2}\tilde{B}^{(0)}Q^{(0)}_{i}, \,\,\,\,
\nonumber \\
& & 
g_{ij}=a^{2}[(1+2H_{L}Q^{(0)})\delta_{ij}+2H_{T}^{(0)}Q_{ij}^{(0)}]                     
\nonumber \\
& &
U_{i} = a v^{(0)} Q^{(0)}_{i}, \,\,\,\, \rho=\rho_{B}(1+\delta Q^{(0)})
\nonumber \\
& &
T_{ij} = P_{B}(1+\pi_{L} Q^{(0)})\delta_{ij} + P_{B}\pi_{T}^{(0)}Q_{ij}^{(0)}  \ ,
\end{eqnarray}     
where function $Q^{(0)} = \exp ({i\vec{k} \cdot \vec{r}})$ is a plane wave, 
$Q^{(0)}_{i} = (-1/k) \partial_{i} Q^{(0)}$, and  $Q_{ij}^{(0)} = k^{-2} \partial_{j}
\partial_{i} Q^{(0)} + (1/3) \delta_{ij} Q^{(0)}$.
The scalar modes $\tilde{A}$, $\tilde{B}^{(0)}$, $H_{L}$, $H_{T}^{(0)}$,
$v^{(0)}$,$\delta$, $\pi_{L}$ and $\pi_{T}^{(0)}$ are functions of $k$ (wavenumber) and $\tau $.
Any other quantity as, e.g., $U_{0}$ and $T_{0i}$, may be easily written in terms 
of these modes. 

The synchronous gauge is fixed by the conditions $\tilde{A}  = \tilde{B}^{(0)} =0$.
In this gauge,  
the modes used in CMBFAST are those defined in Ma \& Bertschinger \cite{mb95}. These modes 
are related to the Bardeen ones as follows:
$h=6H_{L}$, $\eta = -H_{L}- H_{T}^{(0)}/3$, $\theta = k v^{(0)}$, 
$\sigma = 2P_{B}\pi_{T}^{(0)} /3(\rho_{B} + P_{B})$. The same mode $\delta $  
associated to the density contrast is used in paper \cite{bar80} and also
in reference \cite{mb95} and, finally, $\pi_{L} $ is not directly used since it is related to 
$\delta $ by means of the equation of state, e.g., for adiabatic perturbations,
the relation $\pi_{L} = (\rho_{B} / P_{B})
(dP_{B}/d\rho_{B}) \delta$ is satisfied. In addition to the CMBFAST scalar modes, 
a new one is necessary due to the existence of $A^{\mu} $. It is easily verified that 
the mode $\Xi^{(0)} $ defined by 
the equation 
\begin{equation}
\nabla \cdot A = \Xi_{B} (1+ \Xi^{(0)} Q^{(0)}) \ ,
\label{diva_exp}
\end{equation}     
is the most appropriate to write the $A^{\mu} $ field equations in the simplest and 
most operating way. These equations reduces to \citep{dal12}:
\begin{equation}
\ddot{\Xi}^{(0)} + 2 \frac {\dot{a}}{a} \dot{\Xi}^{(0)} + k^{2} \Xi^{(0)}        
= 0  \ .
\label{fe12}
\end{equation}
This second order differential equation does not involve 
the CMBFAST modes associated to the metric and the cosmological fluids.
Apart from the mode $\Xi^{(0)}$, it only involves background functions as  $\dot{a}/a$
and the wavenumber.

Eq.~(\ref{fe12}) is equivalent to the following system of linear 
differential equations 
\begin{equation}
\dot{\Xi}^{(0)}  = \xi^{(0)}
\label{fe12-l1}
\end{equation}    
\begin{equation}
\dot{\xi}^{(0)} = -2 \frac {\dot{a}}{a} \xi^{(0)} - k^{2} \Xi^{(0)} \,
\label{fe12-l2}
\end{equation}    
which have been included in CMBFAST to be solved by using the initial 
values of $\Xi^{(0)}$ and $\xi^{(0)}$ derived below.

In the chosen gauge, Eqs.~(\ref{emtrg}), (\ref{fieles_vt}), and~(\ref{emtee_vt}) lead to the following 
linearized equations for the evolution of the scalar modes $h$ and $\eta $:
\begin{equation}
k^{2} \eta - \frac {1}{2} \frac {\dot{a}}{a} \dot{h} = 4 \pi G 
[-a^{2} \rho_{B} \delta - 2 \gamma \Xi_{B} (a^{2}
\Xi_{B} \Xi^{(0)} + A_{0B} \xi^{(0)} ) ]  
\label{mbl1}
\end{equation}  
\begin{equation}
k^{2} \dot{\eta} = 4 \pi G 
[a^{2} (\rho_{B}+P_{B}) \theta + 2 \gamma k^{2} A_{0B} \Xi_{B} 
\Xi^{(0)}]
\label{mbl2} 
\end{equation}  
\begin{equation}   
\ddot{h} + 2 \frac {\dot{a}}{a} \dot{h} -2k^{2} \eta = 
-24 \pi G 
[a^{2} P_{B} \pi_{L} - 2 \gamma \Xi_{B} (a^{2}
\Xi_{B} \Xi^{(0)} - A_{0B} \xi^{(0)} )]
\label{mbl3} 
\end{equation}  
\begin{equation}
\ddot{h} + 6 \ddot{\eta} + 2 \frac {\dot{a}}{a} (\dot{h} + 6\dot{\eta}) 
-2k^{2} \eta = 
-24 \pi G a^{2} (\rho_{B}+P_{B}) \sigma   \ .
\label{mbl4} 
\end{equation}    
If the terms involving $\gamma $ are canceled, the equations of standard GR cosmology 
labeled (21a)--(21d) in Ma \& Bertschinger \cite{mb95} are recovered. These terms       
--appearing only in VT cosmology-- have been  
included in CMBFAST. Since $A_{0B} $ and $\Xi_{B}$ are proportional to $S_{gn} \gamma^{-1/2} $,
it is obvious that the three VT terms are independent of both $S_{gn} $
and $\gamma $. The values taken by these terms depend 
on the initial values of $\Xi^{(0)} $ and $\xi^{0} = \dot{\Xi}^{(0)} $.

We assume that the universe contains baryons, photons, massless neutrinos, and dark 
matter. The energy momentum tensor of all these components is $T^{\mu \nu}_{GR} $. 
Dark energy is due to the field $A^{\mu} (\tau)$. The background energy
density of this field is constant and its equation of state is 
$W=-1$. There are dark energy 
fluctuations, which have been taken into account to obtain Eqs.~(\ref{mbl1}) --~(\ref{mbl4})
by using the first order approximation of $T^{\mu \nu}_{VT}$.
 
By using Eqs.~(\ref{1.3_vt}) and~(\ref{confic}, it may be easily proved that 
the covariant divergence $\nabla_{\mu} T^{\mu \nu}_{VT} $ vanishes (see also \cite{wil93}).
Hence, according to Eq.~(\ref{fieles_vt}),
the energy-momentum conservation law $\nabla_{\alpha} T^{\alpha \beta}_{GR} = 0$ is satisfied,
as it occurs
in the standard cosmological model based on GR.
Therefore, the variables 
$\delta $, $\theta $, and $\sigma $ corresponding to each 
particle distribution obey the same equations as in standard GR
cosmology and, consequently, we can write (see Eqs.~(92) in paper \cite{mb95}):
\begin{eqnarray}
& &
\dot{\delta}_{\gamma} +\frac {4}{3} \theta_{\gamma} +\frac{2}{3} \dot{h} =0, 
\,\,\,\, \dot{\theta}_{\gamma} - \frac {1}{4} k^{2} \delta_{\gamma} =0, \,\,\,\,  
\nonumber \\
& & 
\dot{\delta}_{\nu} +\frac {4}{3} \theta_{\nu} +\frac{2}{3} \dot{h} =0, 
\,\,\,\, \dot{\theta}_{\nu} - \frac {1}{4} k^{2} (\delta_{\nu} -4 \sigma_{\nu}) =0, \,\,\,\,                   
\nonumber \\
& &                                     
\dot{\sigma}_{\nu} - \frac {2}{15} (2 \theta_{\nu} + \dot{h} + 6 \dot{\eta}) =0 \ ,
\label{arr2}
\end{eqnarray}
where the indices $\gamma $ and $\nu $ make reference to photons and massless neutrinos, 
respectively.
The treatment of the interaction between photons 
and baryons (including reionization) is also identical to that 
described by Ma \& Bertschinger \cite{mb95} and implemented in CMBFAST. 
Finally, from Eqs.~(\ref{mbl1})--~(\ref{mbl4}) and the background field 
equations, one easily 
finds the following differential equation:
\begin{equation}
\tau^{2} \ddot{h} + \tau \dot{h} +6 \delta +32 \pi G \gamma \tau^{2} 
(2 \Xi_{B} A_{0B} \xi^{(0)} - a^{2}
\Xi_{B}^{2} \Xi^{(0)} ) = 0 \ ,
\label{1_92}
\end{equation}
where $\delta = (1-R_{\nu})\delta_{\gamma} + R_{\nu}\delta_{\nu}$,
with $R_{\nu} = \rho_{\nu B}/(\rho_{\nu B}+\rho_{\gamma B})$. Of course, 
this last equation is  satisfied in the radiation dominated era, where 
initial conditions are obtained. It generalizes the first
of Eqs.~(92) in paper \cite{mb95}. We have already found all the equations necessary to fix 
the initial conditions for integrations in VT cosmology. 
Therefore, let us now estimate the initial values of all the scalar modes at 
$z=10^{8} $. Our method to look for these values is similar to that
described in Ma \& Bertschinger \cite{mb95} for the CMBFAST modes, but it has been extended
to take into account the new functions $\Xi^{(0)} $ and $\xi^{0}$. 
It is assumed that, in the radiation dominated era, any mode $X$ may
be expanded in the form
\begin{equation}
X = \sum_{n,m} \beta_{nm} k^{n} \tau^{m} \ ,
\label{gex}
\end{equation}
where the values taken by the integer numbers $n$ and $m$ must be 
fixed for each $X$. The smallness of $k\tau$ for cosmological scales,
the existence of growing and decaying terms in Eq.~(\ref{gex}), and other 
considerations allow us to determine the n and m values being relevant 
for each mode. We begin with
$\Xi^{(0)} $ and $\xi^{(0)} $.

For small enough
scales ($k<<1$), the term proportional to $k^{2}$ in Eq.~(\ref{fe12}) 
may be neglected. Thus, 
this equation reduces to
$\ddot{\Xi}^{(0)} + 2 \frac {\dot{a}}{a} \dot{\Xi}^{(0)} = 0$. The solution of 
this equation is $\xi^{(0)} = \dot{\Xi}^{(0)} \propto a^{-2} \propto \tau^{-2}$.
In order to obtain these last relations, it has been taken into account that
the equation $a=\alpha \tau $ is satisfied in the radiation dominated era.
A new integration leads to 
$\Xi^{(0)} = \tilde{D}_{1} + \tilde{D}_{2} \tau^{-1} $,
where $\tilde{D}_{1} $ and $\tilde{D}_{2} $ are constants of 
integration. 
If the term involving $k^{2} $             
is not neglected, Eq.~(\ref{fe12}) has the following approximating solution:
\begin{equation}
\Xi^{(0)} = D_{1} k^{n} \Big[1 -  \frac {1}{6} k^{2} \tau^{2} + \frac{1}{120} k^{4} \tau^{4} + ... \Big] \ .
\label{ser1}
\end{equation}
which is valid for values of $k\tau $ much smaller than unity.
During a part of the radiation dominated era, including the time 
corresponding to redshift
$z=10^{8} $, all the cosmological scales are outside the effective 
horizon and $k \tau $ is small enough to guarantee the validity of
Eq.~(\ref{ser1}). In this equation, the terms of the form 
$k^{n+2} \tau^{m+2} $ may be neglected against the terms 
$k^{n} \tau^{m} $, which are much greater due to the smallness 
of $k\tau $. Taking into account this fact and Eq.~(\ref{ser1}),
one easily finds the following 
values of $\Xi^{(0)} $ and $\dot{\Xi}^{(0)} $,
\begin{equation}
\Xi^{(0)} = D_{1} k^{n}\ , \,\,\,\,\,\,\,\,\,\,\,\,\,\,\,\,  
\xi^{(0)} = \dot{\Xi}^{(0)} = 0 \ .
\label{ci1}
\end{equation}  
These values correspond to the largest term
of the series (\ref{ser1}) giving $\Xi^{(0)} $. 
Since they do not depend on $\tau $
during the part of the radiation dominated era mentioned above,
the initial values of $\Xi^{(0)}$ and $\xi^{(0)}$, at $z=10^{8} $, 
are $\Xi^{(0)}_{in} = D_{1} k^{n} $ and $\xi^{(0)}_{in} = 0 $.
We see that the initial conditions 
for the scalar modes characteristic of VT only depend on the 
parameter $D_{1} $ and, consequently,
any possible new effect due to cosmological scalar modes 
appearing in VT --but not in GR-- depends on the value of this parameter,
which plays the role of a normalization constant.
Since final results depend on $D_{1} $, 
comparisons with observations should lead to an estimate of this constant.

Let us now look for the initial conditions corresponding to the remaining 
variables to be evolved. 
Our method is analogous to that used in paper \cite{mb95}. 
First of all,  
the terms proportional to $k^{2}$ 
are neglected in Eq.~(\ref{arr2}); thus --as in standard cosmology-- 
the following relations are 
found  $\theta_{\nu} = \theta_{\gamma} = 0 $ and $\delta_{\nu} = \delta_{\gamma} = -(2/3)h $; hence
\begin{equation}
\delta=-(2/3)h 
\label{delh} 
\end{equation} 
and 
\begin{equation}
\theta \equiv (1-R_{\nu})\theta_{\gamma} + R_{\nu}\theta_{\nu} =0 \ .
\label{delthe}
\end{equation}

The second time derivative of Eq.~(\ref{1_92}) with respect to $\tau $ is calculated, and
taking into account Eq.~(\ref{delh}), the following relation is easily obtained:
\begin{equation}
\tau h^{(4)} + 5 h^{(3)} + 32 \pi G \gamma \tau^{-1} \ddot{\mu} = 0 \ ,
\label{dd1}
\end{equation}   
where
\begin{equation}
\mu = \tau^{2} (2 \Xi_{B} A_{0B} \xi^{(0)} - a^{2}
\Xi_{B}^{2} \Xi^{(0)} ) \ .
\label{dd2}
\end{equation}
Taking into account the relations $A_{0B} \tau^{-3}
= (A_{0B})_{in} \tau^{-3}_{in} $, $a\tau^{-1} = a_{in} \tau^{-1}_{in}$, plus  
Eqs.~(\ref{xib}),~(\ref{inib0}), and (\ref{ci1}), quantity 
$\ddot{\mu} $ may be easily calculated and replaced into Eq.~(\ref{dd1}) to get
\begin{equation}
\tau h^{(4)} + 5 h^{(3)} = \frac {384 \pi G D_{1} \rho_{v}} {(1 + z_{in})^{2}}
\Big( \frac {\dot{a}}{a} \Big)_{in} k^{n} \tau 
\label{eor4}
\end{equation}
If the second order derivative of this equation 
--with respect to $\tau $-- is calculated, the following equation is found:
\begin{equation}
\tau h^{(6)} + 7 h^{(5)} = 0 \ ,
\label{orsup}
\end{equation}
where $h^{(6)} $ and $h^{(5)} $ stand for the sixth and fifth order 
derivatives of function $h$ with respect to $\tau $. 
Only the mode $h$ is involved in this equation.
The sixth order differential equation (\ref{orsup}) may be easily integrated.
The solution is a linear combination of the powers $\tau^{-2} $, $\tau^{0} $,
$\tau $, $\tau^{2} $, $\tau^{3} $, and $\tau^{4} $. The powers 
$\tau^{3} $ and $\tau^{4} $ do not appear 
in GR cosmology, where the equation to be solved has the form
$\tau h^{(4)} + 5 h^{(3)} = 0$ (see paper \cite{mb95}).

By using the same arguments as in Ma \& Bertschinger 
\cite{mb95} for the powers $\tau^{-2} $, $\tau^{0} $,
$\tau $, $\tau^{2} $, but taking into account the new dependence in $\tau^{3}$,
and $\tau^{4}$, we write
\begin{equation}
h(k,\tau) = C_{2} (k\tau)^{2} + C_{3} (k\tau)^{3} + C_{4} (k\tau)^{4}\ .
\label{in_h}
\end{equation}
Thus, the initial condition of GR cosmology is recovered for $C_{3}= C_{4}=0 $.
For appropriate values of $C_{3} $ and $C_{4} $, the second and third terms of the right hand side 
of Eq.~(\ref{in_h}) might account for small deviations 
with respect to GR cosmology, which could be compatible with observations. 
By using Eq.~(\ref{in_h}), it is easily seen that, whatever $k$ and $\tau $ may be,
Eq.~(\ref{eor4}) is identically satisfied for $n=4$, 
\begin{equation}
C_{3} = 0\ , \,\,\,\,\,\, C_{4} = \frac {8 \pi G D_{1} \rho_{v}} {3(1 + z_{in})^{2}}
\Big( \frac {\dot{a}}{a} \Big)_{in}^{2} \ .
\label{consts}                          
\end{equation}
Therefore, from Eqs.~(\ref{in_h}) and~(\ref{consts}), it follows that, to lowest order in $k\tau$, 
function $h(k,\tau) $ involves two 
normalization constants $C_{2} $ and $D_{1} $. Constant $C_{2}$ also appears in standard 
GR cosmology, whereas $D_{1} $ is a new independent normalization constant.
Standard cosmology is recovered for $D_{1}= 0$ ($C_{4}=0$). For appropriate $D_{1} $
values, the term $C_{4} (k\tau)^{4}$ may be non negligible and, consequently, it could lead to 
deviations from standard cosmology, which might help to explain  
current observations better.

From Eqs.~(\ref{mbl2}),~(\ref{delthe}), plus Eq.~(\ref{ci1}) with $n=4$, one easily finds 
\begin{equation}
\dot{\eta} = - \frac {8 \pi G D_{1} \rho_{v}} {5(1 + z_{in})^{2}}
\Big( \frac {\dot{a}}{a} \Big)_{in}^{2}  k^{4} \tau^{3} \ .
\label{etap}
\end{equation} 
A simple integration leads to
\begin{equation}
\eta = 2C_{2} - \frac {3} {20} C_{4}  k^{4} \tau^{4} \ .
\label{etap2}
\end{equation} 
Only the first term of the right hand side of this last equation 
arises in standard cosmology. The second term may be neglected 
--to lowest order in $k\tau $-- since 
it involves a very small factor of the form $k^{4} \tau^{4} $. Hence, 
our approximation leads to $\dot{\eta} =0 $. Since $\theta $ also vanishes,
the last of Eqs.~(\ref{arr2}) reduces to  
$\dot{\sigma}_{\nu} = - \frac {2}{15} \dot{h} $.
 
The initial conditions to lowest order in $k\tau $ are summarized
as follows:
\begin{eqnarray}
& &
\delta_{\gamma} = \delta_{\nu} = \frac {4}{3} \delta_{b} = \frac {4}{3} \delta_{c}
= -\frac {2}{3} h, \,\,\,\,   h = C_{2} (k\tau )^{2} + C_{4} (k\tau )^{4},
\nonumber \\
& & 
\theta_{\gamma} = \theta_{\nu} = \theta_{b} = \theta_{c} =0, \,\,\,\, \dot{\eta} = 0, \,\,\,\,  
\dot{\sigma}_{\nu} = - \frac {2}{15} \dot{h} \ ,
\label{arr3}   
\end{eqnarray}
where indices ${b} $ and ${c} $
stand for baryons and cold dark matter, respectively,
and constant $C_{4} $ is given by Eq.~(\ref{consts}). 
The term $C_{4} (k\tau )^{4}$ is not neglected in the formula for $h$ since 
quantity $k^{2}\tau^{2} $ is small, but constant $C_{4}$ may be greater than $C_{2}$.

Let us now combine Eqs.~(\ref{arr2}) --without neglecting the terms involving quantity $k^{2} $--
to go beyond the lowest order in $k\tau $. A lengthy but straightforward calculation 
leads to:
\begin{eqnarray}
& &
\theta_{\gamma} = \theta_{b} = -\frac {1}{18} C_{2} k^{4} \tau^{3}
-\frac {1}{30} C_{4} k^{6} \tau^{5}, \,\,\,\, \theta_{\nu} = - \frac {23+4R_{\nu}}{18(15+4R_{\nu})}
C_{2} k^{4} \tau^{3}-\frac{1}{30} C_{4} k^{6} \tau^{5}
\nonumber \\
& & 
\sigma_{\nu} = \frac {4}{3(15+4R_{\nu})} C_{2} k^{2} \tau^{2}, \,\,\,\, 
\eta = \Big[2- \frac {5+4R_{\nu}} {6(15+4R_{\nu})} (k\tau)^{2} \Big] C_{2} \ .
\label{arr4}
\end{eqnarray}   
Quantities $\delta_{\gamma}$, $\delta_{\nu}$, $\delta_{b}$, $\delta_{c} $, $h$,
and $\theta_{c} $ have the same form as in Eq.~(\ref{arr3}). It is due to the fact that 
the new terms arising beyond the lowest order approximation in $k\tau $ 
are negligible. In the case 
$C_{4} =0$, Eqs.~(\ref{arr4}) reduce to the equations (96) derived by \cite{mb95} 
in the framework of the standard cosmological model. Differences are due to the
terms involving the $C_{4} $ (equivalently $D_{1} $) normalization constant.
 
\section{Numerical results}
\label{sec:4}

All the calculations are performed under the following basic assumptions:
the background is flat, perturbations are adiabatic, the lensing effect
is not considered, there are no massive neutrinos, the equation of state of the dark energy is
$P = W \rho $ with $W=-1$, vector and tensor modes are negligible,
the mean CMB temperature is $T_{CMB}=2.726$, the effective number of relativistic species is 
$3.046$, and the total number of effectively massless degrees of freedom is $g_{*} =10.75 $.

Statistical methods (Markov chains) are used to fit the theoretical
predictions (based on the above basic assumptions) to current observational evidences about
high redshift Ia supernovae (SNe Ia) luminosity and CMB temperature anisotropy. 
In GR (VT), the fit is based on                                                                  
six (seven) parameters. Numerical calculations have been carried out by using modifications of 
the well known codes CMBFAST and COSMOMC. The new codes are hereafter called 
VT-CMBFAST and VT-COSMOMC. These tools have been designed for VT applications.
The code VT-CMBFAST includes the equations and initial conditions obtained in Secs. 
\ref{sec:2} and \ref{sec:3}, which 
are necessary to describe both the VT background and the scalar modes. 
The original CMBFAST code uses the same formalism
as in previous sections, which makes it easy to perform the modifications necessary to include new elements 
characteristic of VT. Since the code CAMB uses other formalism, we have preferred the 
modification of CMBFAST for VT cosmological studies.
Although the original version of COSMOMC 
uses the code CAMB for the numerical estimation of CMB spectra and other quantities, 
we have designed the version VT-COSMOMC 
(for calculations in the framework of VT), which uses VT-CMBFAST instead of CAMB.
 
\begin{table*}
\caption{\label{tab:1} Values of the fitted cosmological parameters.
BF stands for best fit, and the marginalized lower and upper limits of
each parameter, at $2\sigma $ (95 \%) confidence, are listed in the L2 
and U2 cases, respectively.}
\begin{ruledtabular}
\begin{tabular}{llccccccc}
THEORY & CASE & $D_{1}\times 10^{-8}$ & $\Omega_{b}h^{2}$ & 
 $\Omega_{DM}h^{2}$ & $\tau$ & $n_{s}$ & $\log[10^{10}A_{s}]$ & $\theta$ \\ 
\hline
GR & BF & 0.0 & 0.0223 & 0.112 & 0.0836  & 0.962  & 3.067 & 1.039    \\  
GR & L2 & 0.0 & 0.0207 & 0.096 & 0.0460  & 0.920  & 2.967 & 1.030    \\  
GR & U2 & 0.0 & 0.0237 & 0.124 & 0.1285  & 1.000  & 3.168 & 1.047    \\  
VT & BF & 0.203 & 0.0224 & 0.112 & 0.0866  & 0.963  & 3.074 & 1.039  \\
VT & L2 & -5.314 & 0.0189 & 0.082 & 0.0103  & 0.878  & 2.871 & 1.022 \\
VT & U2 & 5.320 & 0.02801 & 0.137 & 0.0203  & 1.119  & 3.324 & 1.054   \\    
\end{tabular}
\end{ruledtabular}
\end{table*}

First of all, with the basic assumptions, the observational data 
and the new codes mentioned in the first paragraphs of this section,
we have found the best fit in the framework of GR ($D_{1} =0$). The six parameters 
used to fit predictions and observations are
$\Omega_{b}h^{2}$, $\Omega_{DM}h^{2} $, $\tau $, $n_{s} $,
$\log[10^{10}A_{s}]$, and $\theta $, where 
$\Omega_{b}$ and $\Omega_{DM}$ 
are the density parameters of baryons and dark matter, respectively,
$h$ is the reduced Hubble constant, $\tau $ is the reionization optical depth, 
$n_{s} $ is the spectral index of the power spectrum of scalar modes, 
and $A_{s} $ is the normalization constant of the same spectrum whose form is 
$P(k) = A_{s} k^{n_{s}} $, finally, the parameter $\theta $ is defined by 
the relation
$\theta \times 10^{-2}= d_{A}(z_{*})/r_{s}(z_{*})$, where 
$d_{A}(z_{*}) $ is the angular diameter distance at decoupling redshift $z_{*} $,
and $r_{s}(z_{*})$ is 
the sound horizon at the same redshift.
The resulting values of the above six parameters 
corresponding to our best fit in GR are given in the 
first row of Table \ref{tab:1}. These values are compatible with 
those of Table 8 in \cite{jar11}, which were obtained from the
Wilkinson microwave anisotropy probe seven (WMAP7) years data.
For each parameter, the 
second (third) row of this Table defines the lower (upper) limit
of an interval, which contains the true value of the chosen parameter,
at $95$ \% confidence, in the marginalized case; namely, 
if the remaining five parameters are  
chosen to be those of the first row of Table \ref{tab:1}(best fit).

\begin{figure}[tbh]
\includegraphics[angle=0,width=0.8\textwidth]{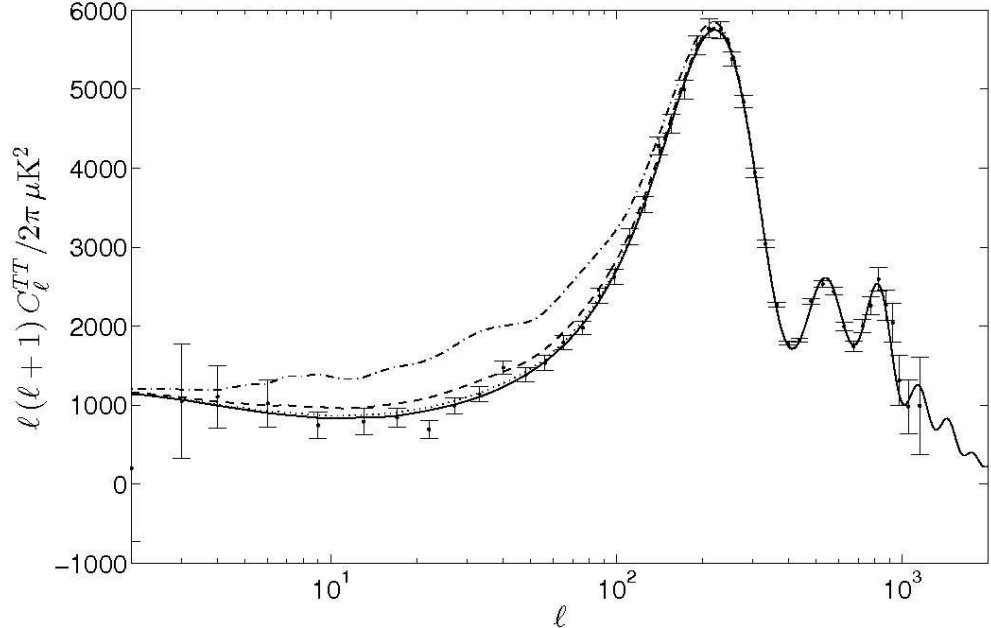}%
\caption{\label{fig1} CMB temperature angular power spectra in terms of $\ell$.   
Solid, dotted, dashed, and dotted-dashed lines correspond to $D_{1}=0$,
$D_{1}=2\times 10^{8}$, $D_{1}=4\times 10^{8}$, and $D_{1}=8\times 10^{8}$, respectively.
Dots with error bars are WMAP7 measurements.}
\end{figure}

We have used VT--CMBFAST to find the 
following CMB angular power spectra:
(i) the $C^{TT}_{\ell} $  ($C^{EE}_{\ell} $) coefficients measuring CMB temperature
(E-polarization) correlations at angular scales $\alpha = \pi / \ell $
with $\ell < 2100 $, 
and (ii) the parameters $C^{ET}_{\ell} $ 
giving the cross correlations between temperature and E-polarization for the 
same scales. The resulting $C^{TT}_{\ell} $ quantities
corresponding to various cases are presented in Fig.~\ref{fig1}. In all these cases,  
the values of the six parameters used in our previous fit (first row of Table    
\ref{tab:1}) have been fixed, whereas parameter $D_{1} $ has been varied.
For $D_{1} =0$ (solid line) the angular power spectrum corresponds to our
GR best fit (first row of Table \ref{tab:1}).

As it follows from Fig.~\ref{fig1}, 
for $D_{1} = 2 \times 10^{8} $ (dotted line), the resulting angular power spectrum is very similar
to that obtained for $D_{1} = 0 $ (solid line). Moreover,
from the shape
of the dotted, dashed ($D_{1} = 4 \times 10^{8} $), and dotted-dashed 
($D_{1} = 8 \times 10^{8} $) lines, it follows that the deviations with respect to the solid line
(effect due to $D_{1} $) increase as $|D_{1}| $ grows. 
For some $\ell $ values, the dotted-dashed line deviates too much 
from the solid line, which corresponds to $D_{1} =0$.
In the same figure we also see that, for all the $D_{1} $ values, 
the deviations with respect to the solid line are: (i) negligible 
for $\ell $ values greater than $ \sim 250 $, which means that only the angular scales greater than 
$\sim 0.72 $ degrees are significantly affected by the VT scalar mode $\Xi^{(0)} $ and, (ii)
small for $\ell < 4$.

Moreover, by using the VT--CMBFAST code, we have verified that: 
($\alpha $) the deviations with respect to the solid line of Fig.~\ref{fig1} do not depend on the 
sign of $D_{1} $, but only on $|D_{1}|$, and ($\beta $) for the three non vanishing 
$D_{1} $ values considered in Fig.~\ref{fig1}, the $C^{ET}_{\ell} $ and $C^{EE}_{\ell}$ spectra are 
indistinguishable from those corresponding to $D_{1} =0$.

\begin{figure}[tbh]
\includegraphics[angle=0,width=0.8\textwidth]{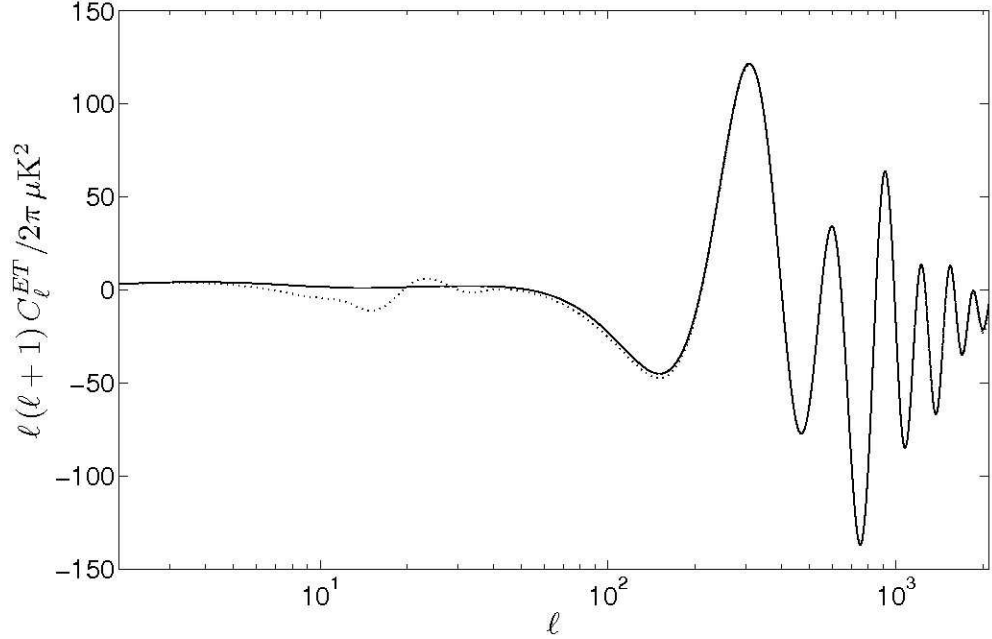}%
\caption{\label{fig2} CMB angular power spectra for the ET cross correlation in terms of $\ell$.
Solid and dotted lines show the resulting spectra for $D_{1}=0$ and $D_{1}= 10^{11}$,
respectively.}
\end{figure}

The $C^{ET}_{\ell} $ spectrum corresponding to $D_{1} = 10^{11}$ is represented 
in the dotted line of Fig.~\ref{fig2}. This line slightly deviates with respect to 
the solid one, which has been obtained for $D_{1} =0$. 
These small $C^{ET}_{\ell} $ deviations are visible for $\ell $ smaller than $\sim 200 $. 
Nevertheless, for $D_{1} = 10^{11}$,
the $C^{TT}_{\ell} $ spectrum would be too different from that shown in the solid line 
of Fig.~\ref{fig1} and, consequently, this high $D_{1}$ value is not  
admissible. The 
same occurs with the $C^{EE}_{\ell}$ spectrum, which begins to be different from 
that of the case $D_{1} =0$ for values as great as $D_{1} \simeq 10^{13}$.
All this means that VT--CMBFAST rightly estimates the $C^{ET}_{\ell} $ and $C^{EE}_{\ell} $
coefficients, but they are negligible for any realistic $|D_{1}| $ value smaller 
than $ 8\times 10^{8} $ (dotted-dashed line of Fig.~\ref{fig1}).

\begin{figure}[tbh]
\includegraphics[angle=0,width=0.8\textwidth]{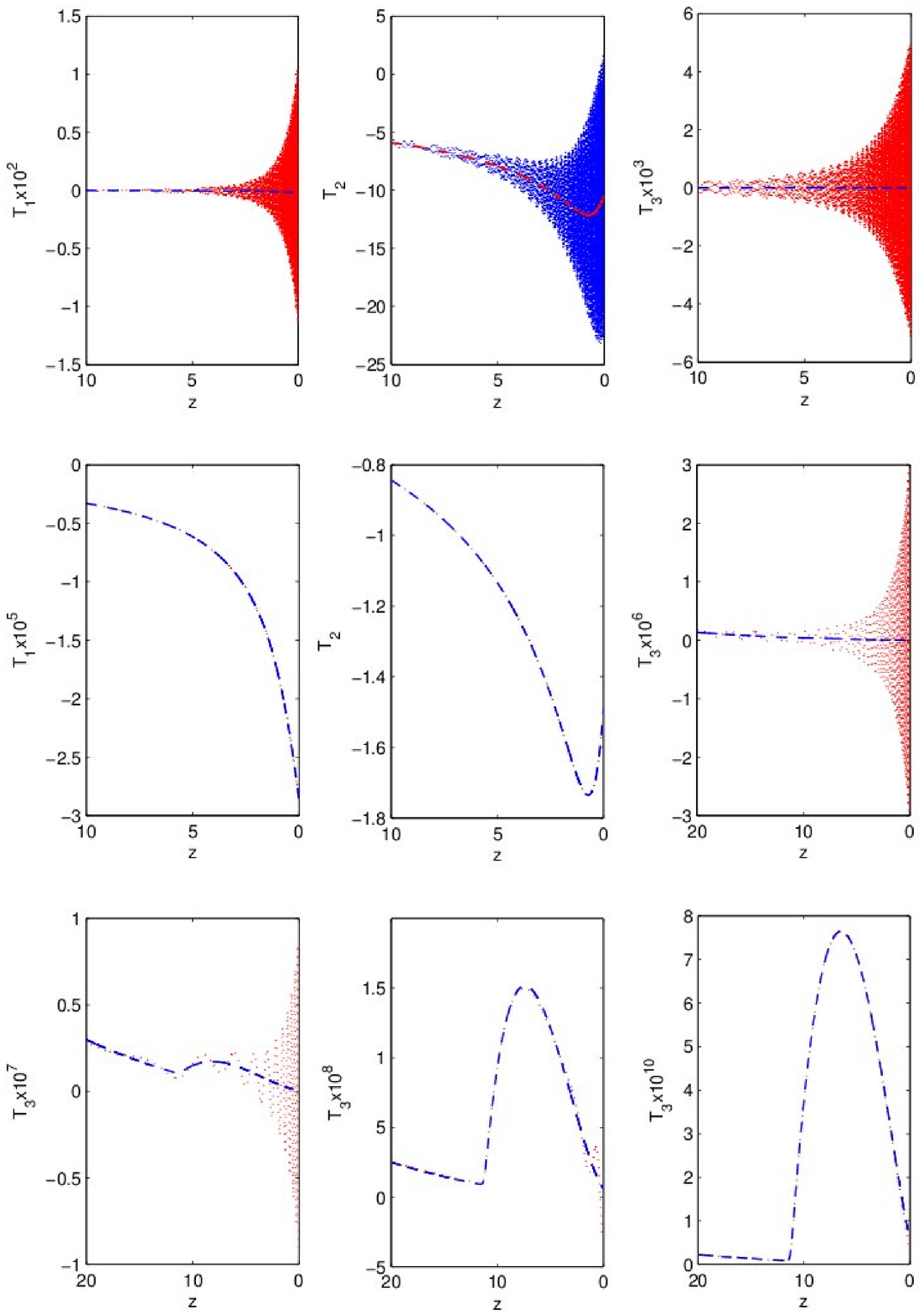}%
\caption{\label{fig3} Top and middle panels: representation of functions $T_{1} = a \ddot{h} + \dot{a} \dot{h} $ (left), 
$T_{2} = \dot{h} $ (central), and $T_{3} = \dot{\eta} $ (right) in terms of the redshift $z$ for various scales of interest.
Top (middle) panels correspond to a spatial scale close to $16h^{-1} \ Mpc$ ($200h^{-1} \ Mpc$).
Bottom panels: function $T_{3} $ for scales of $700h^{-1} \ Mpc$ (left), $2800h^{-1} \ Mpc$ (central),
and $31400h^{-1} \ Mpc$ (right). In all panels the dotted (dashed) lines show the functions of VT (GR),
except in the top central panel, where the dotted (dashed) lines correspond to GR (VT).}
\end{figure}

In order to understand some of the above results, it is worthwhile to show 
some outputs given by VT--CMBFAST. In Fig.~\ref{fig3}, these outputs are represented 
--in terms of the redshift $z$-- for appropriate spatial scales. 
The chosen outputs are the following functions of $z$:
$T_{1} = a \ddot{h} + \dot{a} \dot{h} $, $T_{2} = \dot{h} $, and 
$T_{3} = \dot{\eta} $, which have been chosen by the following reasons:
(a) they are 
involved in the equations describing the evolution of the 
photon distribution function [see Eq.~(63) in \cite{mb95}], 
which are used to calculate the CMB angular power spectra, and (b) 
they depend on time derivatives of the metric perturbations $\eta $ and 
$h$, whose VT and GR values start to be different at some 
redshift which must be estimated (see \ref{sec:1}).

Functions $T_{1} $,
$T_{2} $, and $T_{3} $ have been obtained for the code runs 
leading to the solid and dotted-dashed   
spectra of Fig.~\ref{fig1}, which correspond to $D_{1}=0$ (GR) and $D_{1}=8\times 10^{8}$
(VT), respectively. 
In the top panel of Fig.~(\ref{fig3}), the 
spatial scale is $L \simeq 16h^{-1} \ Mpc$ (used to define the standard 
parameter $\sigma_{8}$). In the top left ($T_{1}$) and top right ($T_{3}$) panels, 
the blue dashed lines correspond to GR, whereas the red 
dotted ones show the outputs in VT for the chosen $D_{1}$ value. The 
dotted lines (VT) oscillate around the dashed ones (GR). In the top central
($T_{2}$) panel, the red dashed lines correspond to VT, 
whereas the blue dotted ones show the outputs in GR. The 
dotted line (GR) oscillates  around the dashed one (VT). In all cases we find 
oscillations. Quantities $T_{1} $ and $T_{3} $ oscillate in VT, but not in GR, 
whereas $T_{2} $ undergoes oscillations in GR, but not in VT.
In all the middle panels, the spatial scale is $L \simeq 200h^{-1} \ Mpc$.
In these panels the blue dashed lines have been obtained for GR,
and the red dotted ones correspond to VT. By comparing 
the middle panels with the top ones one easily see that, as the spatial scale grows, 
the functions $T_{1}$ and $T_{2} $ obtained in
GR and VT tend to the same limit. For the spatial scale $L \simeq 200h^{-1} \ Mpc$,
the dotted and dashed lines of the 
middle left and central panels are indistinguishable; 
however, for the same scale,  
the $T_{3} $ function corresponding to VT oscillates around its GR values (middle right panel).
In the bottom panels, the spatial scale is varied to see the behavior of the $T_{3} $
function.
The spatial scales increase from left to right taking 
on the values $L \simeq 700h^{-1} \ Mpc$ (left), $L \simeq 2800h^{-1} \ Mpc$ (central),
and $L \simeq 31400h^{-1} \ Mpc$ (right). As it follows from these panels, the oscillations 
of function $T_{3}$ decrease as the spatial scale increases, which means that the VT and GR 
values of $T_{3} $ converge as the spatial scale grows. We see that, for scales larger than 
$\sim 2800h^{-1} \ Mpc$ there are no significant differences between the VT and GR
values of $T_{3} $ (see the bottom central and right panels).

The oscillatory character of the 
differences between GR and VT explains the fact that the VT spectra do not depend 
on the sign of $D_{1} $, but only on $|D_{1}| $. Equivalent oscillations arise for
both $D_{1} $ and $-D_{1} $. Moreover, from Fig.~\ref{fig3} it follows that, 
if there are oscillatory differences for a certain spatial scale,  
they are visible for redshifts smaller than $\sim 5$.
These redshifts are significantly larger than $0.8 $, which is very close
to the redshift corresponding to the beginning of the accelerated expansion 
($2\rho_{v}=\rho_{matter}$) in the standard concordance model. 
Moreover, there are no visible oscillatory differences for 
very large spatial scales, which qualitatively explains why the GR and VT angular power
spectra of Figs.~\ref{fig1} and~\ref{fig2} are more and more similar as 
$\ell $ decreases from $\ell \sim 5 $. 

Figure~\ref{fig1} suggests that VT may explain the observational data for
some non vanishing $D_{1} $ values combined with appropriate 
values of the remaining parameters. In order to verify this suspicion,
let us use a set of parameters to fit 
appropriate observational data and VT predictions by means of statistical techniques. 
The code VT-COSMOMC has been used to perform this fit. We have 
used the seven parameters 
of Table \ref{tab:1}. Only data relative to 
SNe Ia and CMB anisotropy observations have been taken into account. This
choice seems to be appropriate, since the same data lead to very good fits 
in the standard GR model. 
The CMB angular power spectra used by VT-COSMOMC were obtained from 
the WMAP7 data. The last version 
of COSMOMC uses data from PLANCK and WMAP9; nevertheless, 
this version was delivered very recently, after the numerical calculations presented 
in this paper --which are good enough-- were finished.
Further research based on PLANCK spectra will be developed in 
future (see Sec. 5).

\begin{figure}[tbh]
\includegraphics[angle=0,width=0.7\textwidth]{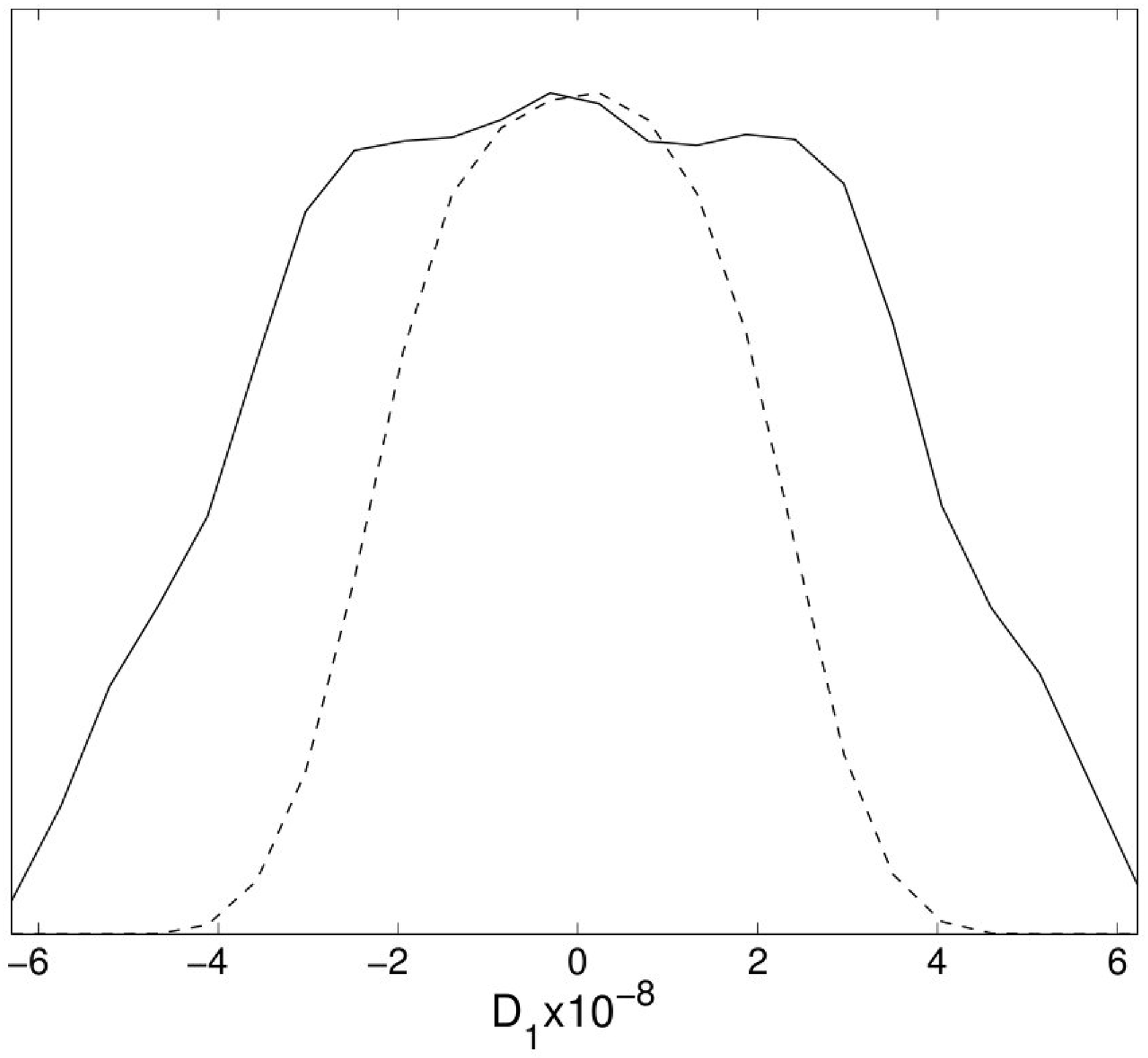}%
\caption{\label{fig4} Solid (dashed) line shows the mean (marginalized) likelihood function 
for the parameter $D_{1} $ of VT.}
\end{figure}

\begin{figure}[tbh]
\includegraphics[angle=0,width=0.9\textwidth]{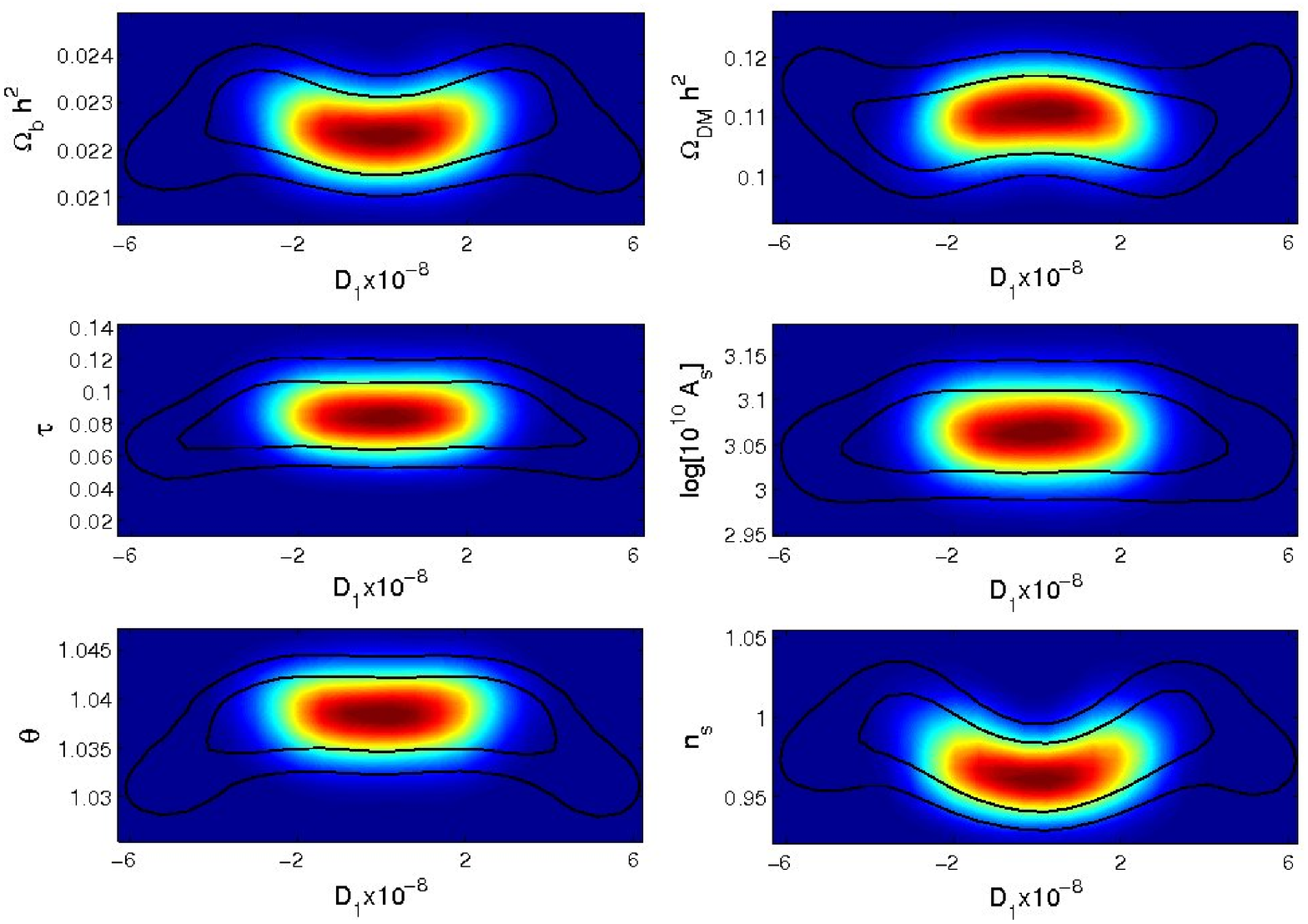}%
\caption{\label{fig5} Each panel corresponds to a pair of parameters. In all panels, one of these parameters is 
$D_{1}$. The second parameter is one of the six GR parameters 
listed in Table \ref{tab:1}.
The grayscale (red-yellow-blue) central region of each panel represents the mean likelihood function. 
The external (internal) contour shows the $2\sigma $  ($1\sigma $) confidence level
in the marginalized case.}
\end{figure}

Results obtained with VT-COSMOMC are presented in Figs.~\ref{fig4} and~\ref{fig5}
and also in Table \ref{tab:1}. Let us discuss the most significant aspects 
of these results. The best fit in VT
corresponds to the parameters shown in the fourth row of Table \ref{tab:1}, 
where we see that $D_{1} $ is very close to  $2\times 10^{7} $ and the remaining parameters
take on values very similar to those of the first row (best fit in standard GR cosmology);
hence, in a representation as that of Fig.~\ref{fig1}, the angular power spectra of the VT and GR 
best fits 
would be indistinguishable. From the point of view of the best fits, both theories are
equivalent, which is a good result for a theory as VT, which explains the existence 
of the cosmological constant.
Nevertheless, let us now show that a more exhaustive statistical analysis strongly
suggests that VT cosmological models 
may be preferable.

The dashed (solid) line of Fig.~\ref{fig4} shows the marginalized (mean) 
likelihood function (with arbitrary normalization) for the analyzed samples of parameters. 
In the marginalized case, the six parameters of the GR models are fixed and their values
are taken to be identical to those of the best VT fit. 
Although the dashed line has a maximum for $D_{1} \simeq 2\times 10^{7} $,
this curve is rather flat around the maximum and it may be stated that values of $D_{1} $ satisfying the 
relation $-10^{8} < D_{1} < 10^{8} $ are also very likely in order to explain the observation. 
A broader interval of admissible $D_{1} $ values is found from the solid line (mean likelihood) of 
Fig.~\ref{fig4}. This line has a wide plateau around the maximum at $D_{1} \simeq 2\times 10^{7} $, 
which means that, if the seven parameters are varied (no marginalization), the mean likelihood 
function takes on values similar to the maximum one for the 
$D_{1}$ values of the plateau and, moreover, for any of these $D_{1} $ values,
there must be likelihood values greater than the mean one, which 
must be closer to the maximum likelihood (see Fig.~\ref{fig4}). 
A visual analysis of this figure shows that 
the plateau is approximately defined by the condition 
$-3 \times 10^{8} < D_{1} < 3 \times 10^{8} $. For these values and appropriate 
values of the six GR parameters, which will be different from those of the best fit, the 
observations may be explained with high probability. 

More statistical information may be found in Fig.~\ref{fig5}, where each panel 
shows the likelihood function for a pair of parameters. One of them is always 
$D_{1} $ and the second one is another of the parameters of Table \ref{tab:1}. 
The grayscale (red-yellow-blue) central zone shows the mean likelihood of the 
considered parameter samples. In all cases, the $D_{1} $ values of this zone 
approximately satisfy the relation $-3 \times 10^{8} < D_{1} < 3 \times 10^{8} $,
in agreement with the discussion of the last paragraph. 
The internal (external) contour shows the 68\% (95\%) confidence limit
in the marginalized case, in which, the remaining five parameters are fixed according to the
best VT fit (fourth row of Table \ref{tab:1}). The external contour tell us that, 
inside the seven  
intervals (one for each parameter) defined by the lower and upper limits given in the two last 
rows of Table \ref{tab:1}, there are values of the seven parameters 
explaining the observational data at $2\sigma $ confidence. In particular,  
the $D_{1}$ value will be in the interval ($-5.3 \times 10^{8}, \, 5.3 \times 10^{8}$). 
According to Fig.~\ref{fig1}, for $|D_{1}| \leq 3\times 10^{8}$,
the VT and GR angular power spectra of the CMB are slightly different for $\ell \leq 250$, and
these spectra are  
rather different for $|D_{1}| \leq 5.3\times 10^{8}$. 
All these considerations indicate that 
there are good fits
for a wide interval of $|D_{1}|$ values. This fact seems to be related 
to: (a) the cosmic variance, which is important for the 
range of $\ell $ values ($\ell < 250$) affected by the condition $D_{1} \neq 0$,
and (b) the CMB spectrum, which remains unchanged for $\ell > 250$ whatever the  
$D_{1} $ value maybe. 
We may also verify that, for the remaining six parameters, 
the VT intervals defined by the lower and upper $2\sigma $ limits given in
the two last rows of Table \ref{tab:1} are wider than the corresponding intervals 
of the GR model ($D_{1} = 0$), whose lower and upper $2\sigma $ limits are 
shown in the second and third rows of the same Table. 
All these considerations indicate  that the seven parameters are coupled and, consequently, 
parameter $D_{1} $ plays an important role in VT statistical fits.

\section{Discussion and conclusions}
\label{sec:5}

It has been proved (see Sec. \ref{sec:2}) that, in EE and VT, the 
background energy density 
of the field $A^{\mu} $
plays the role of dark energy with $W=-1$; nevertheless, 
in order to have a positive dark energy, the coupling 
constant $\gamma $ must be positive (negative) in VT (EE).

In Eqs.~(\ref{emtee}) and~(\ref{emtee_vt}), we see that the last
terms of the right hand side have the same form but opposite signs.
This fact has been justified with a detailed variational study.
Since only these terms contribute to $\rho^{A}_{B}$ ($F^{\mu \nu} = 0$ 
in the background), this density appears to have opposite 
signs in Eqs.~(\ref{eqest}) and~(\ref{eqest_vt}), which 
correspond to EE and VT, respectively. From these 
equations and the condition $\rho^{A}_{B}>0$, the sign 
of $\gamma $ is fixed in both theories.

Since the conservation equations~(\ref{conlaw}) have played a 
very relevant role in the Lagrangian formulation of EE,
a few words about the conserved currents of VT and EE
are worthwhile.
As it follows from Eq.~(\ref{1.3}),           
the conserved current of EE is 
$J_{\mu} + J^{^{A}}_{\mu}$ [see Eq.~(\ref{conlaw})]. 
In the case $J^{\mu} =0$ (VT), the Lagrangian 
$- \frac {1}{4} F^{\mu \nu } F_{\mu \nu }
+\gamma (\nabla_\mu A^{\mu})^{2}$ is invariant under the 
local gauge transformations $A^{\prime \mu} =A^{\mu} + \nabla^{\mu} \Phi$, 
with $\nabla_{\mu} \nabla^{\mu} \Phi =0$ and, consequently, the second
Noether theorem may be applied to 
get the conserved current $J^{^{A}}_{\mu} $ [see Eq.~(\ref{confic}]. 
For $J^{\mu} \neq 0$ (EE), the Lagrangian is
$- \frac {1}{4} F^{\mu \nu } F_{\mu \nu }
+\gamma (\nabla_\mu A^{\mu})^{2} + J^{\mu}A_{\mu}$. It may be 
easily proved that 
this Lagrangian is also gauge invariant, under the above local 
gauge transformation, if $J^{\mu} $ is replaced by 
$\nabla^{\nu} F_{\mu \nu} - J^{^{A}}_{\mu}$, namely,
if $J^{\mu} $ is constrained to satisfy the field equations 
(\ref{1.3}). 
From the resulting gauge invariant Lagrangian
and the second Noether theorem, it follows that the conserved 
current is $J_{\mu} + J^{^{A}}_{\mu}$.
In general, currents $J_{\mu} $ 
and $J^{^{A}}_{\mu}$ are not expected to be separately conserved,
since we should not have two independent conserved currents 
associated to an unique group of local gauge 
transformations. 
 
We have verified that, in a neutral universe where the background current $J^{\mu} $
and its scalar perturbations vanish \citep{dal12}, 
EE and VT lead to the same cosmological conclusions in the study of both the 
background universe and the scalar perturbations; nevertheless,  
as a result of the negative $\gamma $ value involved in EE, which would lead to 
problems with quantification, our results are presented in the 
framework of VT. 
This theory is based on action (\ref{VT.2}), which 
has four terms. 
Deviations with respect to GR only can be produced by the second 
and third terms,
which vanish for $\gamma=\varepsilon =0 $. In other words, for vanishing $\gamma $
and $\varepsilon $, 
action \ref{VT.2} reduces to the GR one and, consequently, for small enough 
values of $\gamma $ and $\varepsilon $, VT and GR would be indistinguishable.

According to Eq.~(\ref{xib}), parameter $\gamma $ must satisfy the
relation $\gamma (\nabla \cdot A)_{B}^{2} = \rho_{v} $. Furthermore, 
as it has been shown in previous sections (see also paper \cite{dal12}), 
there are no additional cosmological 
constraints to be satisfied by the constant quantities $\gamma $ and $(\nabla \cdot A)_{B} $.
It is due to the fact that these quantities may be  
eliminated 
from the evolution equations of the scalar perturbations. Moreover, these equations do 
not involve the parameter $\varepsilon $ either. 
This means that, in cosmology, quantities $\gamma $ and $\varepsilon $ 
only must satisfy the inequality $2\varepsilon - \gamma > 0$.

The strength of gravitation is fixed 
by the first term of action \ref{VT.2} (proportional to R).
The second and third terms --related to gravitation in VT-- should involve small coupling constants
compatible with the weak character of the gravitational interaction;
namely, these constant must be compatible with the fact that 
the strength of the gravitational field is 
very low as compared to the strengths of electroweak 
and strong interactions. Appropriate values of the 
free constants $\gamma $ and 
$\varepsilon$ --which have not been fixed by cosmological considerations-- 
may be chosen (with the constraint $2\varepsilon - \gamma > 0$) to guaranty that the
second and third terms of action \ref{VT.2} 
have nothing to do with strong and electroweak interactions, but with 
gravity.

A general formalism to evolve the VT scalar modes from the redshift 
$z=10^{8} $ is developed. The evolution equations and the initial conditions for all the scalar modes
are written in momentum space (Bardeen formalism) by using the synchronous gauge. Moreover, 
the scalar mode associated to the VT field $A^{\mu} $
is chosen in such a way that:
(i) it evolves separately and, (ii) it is involved in the evolution equations 
for the scalar modes of GR cosmology (standard model). Our methodology is 
analogous to that used by Ma \& Bertschinger \cite{mb95}. Equations and initial conditions are 
fully general.

Our calculations with VT-CMBFAST prove that some time derivatives of the metric modes 
$\eta $ and $h$ (which are involved in the 
evolution equations of the CMB photon distribution function) evolve in the same way 
--in both GR and VT-- until redshifts $\sim 10 $; then, the evolution of these derivatives 
starts to be different in both theories and, at redshifts $z \leq 5 $, 
they take on fully different values in VT and GR, except for very large spatial scales
(see Sec. \ref{sec:4}).
Deviations between VT and GR are oscillatory. They explain the differences between the CMB angular power spectra 
of both theories for $\ell \leq 250 $. 

By using the code VT-COSMOMC, WMAP7 and SNe Ia data have been 
adjusted to VT predictions by using seven parameters. In the standard GR model, 
either WMAP7 or WMAP9 and 
other data (supernovae, matter power spectrum and so on) 
are well fitted with a minimal model involving six parameters (see \cite{jar11,hin12}).
We add the new parameter $D_{1}$ which is characteristic of VT to perform a fit based on 
seven parameters. In the best fit, the six common parameters of the GR and VT models are very similar,
which means that VT works as well as GR; however, there are also good fits for $D_{1}$ values
satisfying the condition $|D_{1}| < 3\times 10^{8} $  and, moreover, at 95\% confidence, 
the parameter $D_{1}$ satisfies the condition $|D_{1}| < 5.3\times 10^{8} $ (see Sec. \ref{sec:4}). 
The fact that we have found good fits for a wide range of $D_{1} $ values
strongly suggests that VT models may explain cosmological observations better than GR. It is due to
the existence of an additional degree of freedom (parameter $D_{1}$), which has a good behavior 
and helps us to get good fits.

A new version of COSMOMC has been recently delivered. It includes PLANCK CMB spectra. 
We are trying to modify this version for 
future applications to VT. New fits based on the modified code 
would use better observational data and, moreover, these fits could 
involve more parameters, 
lensing, and other effects; nevertheless, the study of these general fits is
beyond the paper scope. Here, 
we essentially point out that VT deserves attention, since it is a theory which 
explains: the existence of a cosmological constant, and recent 
CMB and SNe Ia observations (with a minimal model involving seven parameters).
Moreover, parameter $D_{1} $ seems to be a help to fit predictions and observations
in VT and, consequently, VT fits seem to be more promising than the GR ones.

\begin{acknowledgments}
This work has been supported by the Spanish
Ministry of {\em Econom\'{\i}a y Competitividad},
MICINN-FEDER project FIS2012-33582 and CONSOLIDER-INGENIO project 
CSD2010-0064.
We thank Javier Morales (Universidad Miguel Hern\'{a}ndez) for 
comments and suggestions about statistics. 
\end{acknowledgments}

\bibliography{apssamp}

\end{document}